\documentclass[twocolumn,showpacs,english,amsmath,amssymb, pra, aps, 10pt] {revtex4-1}

\usepackage{tgtermes}
\usepackage[utf8]{inputenc}
\usepackage{graphicx}
\usepackage{dcolumn}
\usepackage{bm}
\usepackage{xcolor}
\usepackage{makeidx}
\usepackage{amsfonts}
\usepackage{subfigure}
\usepackage{epsfig}
\usepackage{pst-grad} % For gradients
\usepackage{pst-plot} % For axes
\usepackage[bookmarks=false,colorlinks=true, breaklinks=true]{hyperref}
\hypersetup
	{ colorlinks,%
		citecolor=green,%
		linkcolor=red,%
		filecolor=magenta,%
		urlcolor = cyan,% it can change the color of hyperref in reference
	}
\newcommand{\nn}{\nonumber\\}

\newcommand{\bea}{\begin{eqnarray}}
\newcommand{\ea}{\end{eqnarray}}
\newcommand{\eea}{\end{eqnarray}}

\newcommand{\bfk}{{\bf k}}

\newcommand{\bfq}{{\bf q}}
\newcommand{\sumint}[1]

\begin{document}

\title{Nonadiabatic multichannel dynamics of a spin-orbit-coupled condensate}

\author{Bo Xiong,$^{1,2}$ Jun-hui Zheng,$^{1}$ and Daw-wei Wang$^{1,3}$}
\affiliation{$^{1}$Department of Physics, National Tsing-Hua University, Hsinchu, Taiwan 300, Republic of China \\
             $^{2}$Institute for Advanced Study, Tsinghua University, Beijing 100084, China \\
             $^{3}$Institute of Physics, National Center for Theoretical Science, Hsingchu, Taiwan 300, Republic of China}

\date{\today}

\begin{abstract}
   We investigate the nonadiabatic dynamics of a driven spin-orbit-coupled Bose-Einstein condensate in both weak and strong driven force. It is shown that the standard Landau-Zener (LZ) tunneling fails in the regime of weak driven force and/or strong spin-orbital coupling, where the full nonadiabatic dynamics requires a new mechanism through multichannel effects. Beyond the semiclassical approach, our numerical and analytical results show an oscillating power-law decay in the quantum limit, different from the exponential decay in the semiclassical limit of the LZ effect. Furthermore, the condensate density profile is found to be dynamically fragmented by the multichannel effects and enhanced by interaction effects. Our work therefore provides a complete picture to understand the nonadiabatic dynamics of a spin-orbit coupled condensate, including various range of driven force and interaction effects through multichannel interference. 
The experimental indication of these nonadiabatic dynamics is also discussed.
\end{abstract}


\maketitle
%%%%%%%%%%%%%%%%%%%%%%%%%%%%%%%
\section{Introduction}  

The experimental realization of the synthetic gauge field \cite{lin2, lin3, lin4} and spin-orbit (SO) coupling \cite{lin1} in neutral quantum gases paves the way for studying some exotic many-body physics, including the spin-Hall effect \cite{sino, kane}, Majorana fermions \cite{sau}, and topological insulators \cite{hasa, gali}. In SO-coupled bosonic gas, many theoretical works are involved in the investigation of the ground state \cite{stan, wang, zhou, li}, quantum and thermal fluctuation \cite{ozaw}, collective excitation \cite{li1}, and finite-temperature properties \cite{ozaw1, zhen}. Recently, the condensate dynamics have been experimentally investigated in the presence of SO coupling \cite{olso}, which can be flexibly controlled in a broad parameter range by the external field. This provides an opportunity to investigate the nonadiabatic dynamics, which may not easily be controlled or explored in the conventional solid-state system. 

Nonadiabatic dynamics of an interacting many-body system are theoretically nontrivial and highly challenging because analytic and numerical difficulties are, in general, manifested and the associated results are conventionally nonuniversal. Some theoretical works have attempted to solve such issues, e.g., universal adiabatic dynamics near the regime of the quantum critical point \cite{polk1}, quench-induced phase transition in the quantum Ising model \cite{zure} and in the two-level bosonic system \cite{bo}, and nonadiabaticity induced by fluctuations in a driven Landau-Zener system \cite{altl}. By far, the physically most promisingly realizable and controllable setup to explore nonadiabatic dynamics is the system of a Bose-Einstein condensate (BEC), which has a macroscopic occupation in a single particle state. As a generic example, a BEC is driven by an external field to transport between different energy bands or levels; such nonadiabatic dynamics can be understood using the celebrated Landau-Zener (LZ) mechanism as a starting point \cite{land, zene, stuc, majo}, for example, the interband tunneling in a tilted optical lattice \cite{wu, witt, zene1}. Recent experiments \cite{olso} show that the nonadiabatic dynamics of a \emph{noninteracting} SO-coupled BEC driven by a \emph{finitely large} force can be explained by LZ tunneling. Taking the weak driven force and interatomic interaction into account in this SO-coupling system, the transport dynamics can be, however, much more complicated because the tunneling rates for particles in different momentum channels are distinct. Therefore, some fundamental questions in such system have arisen and need to be answered immediately, in particular, 
how the quantum interference between different momentum channels competes with single-channel LZ dynamics and how the two-body interaction acts on the nonadiabatic dynamics.

In this paper, we investigate the nonadiabatic dynamics of a SO-coupled condensate driven through the avoided crossing point opened by spin-orbital coupling [Fig.\,\ref{model}(b)], including the multichannel effects in momentum space. We find that the full dynamics is controlled by the relative amplitude of the initial driven kinetic energy compared to the strength of spin-orbital coupling: When the kinetic energy dominates, the condensate dynamics can be approximated by a semiclassical ``particle'' in momentum space, leading to the single-channel Landau-Zener tunneling. However, in the regime of strong spin-orbital coupling, the nonadiabatic dynamics is dominated by the quantum interference effects through multichannel coherent tunneling in momentum space, leading to a universal oscillating power-law decay. We further find a dynamical fragmentation of the condensate when the nonadiabatic dynamics evolves for a longer time, which also results from the SO coupling and enhanced by interparticle interaction. All of these novel features of nonadiabatic dynamics can be observed in the current experimental condition. 
                            
This paper is organized as follows. In Sec.\,\ref{secII}, we will first introduce the underlying many-body system Hamiltonian for a SO-coupled condensate. Analytic studies and numerical results for the nonadiabatic tunneling dynamics in the noninteracting regime by multichannel interference are shown in Sec.\,\ref{quantum_limit}, while the results in the semiclassical limit are in Sec.\,\ref{classical_limit}. In Sec.\,\ref{secIV} we show the interaction effects on the tunneling dynamics and we summarize our results in Sec.\,\ref{secV}. 

%%%%%%%%%%%%%%%%%%%%%%%%%%%%%%%%%%%%%%%%%%%%%%%%%%
\section{Model and Hamiltonian} 
\label{secII}
\begin{figure}[t]
\centering
\includegraphics[scale = 0.367]{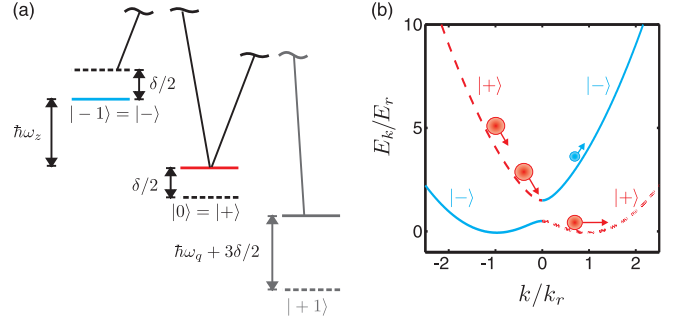} 		        	     
\caption{(Color online) (a) Level diagram for SO-coupled system. Two spin states, $|F = 1, m_{F} = 0 \rangle = |+ \rangle$ and $|F = 1, m_{F} = -1 \rangle = |- \rangle$, are coupled by two lasers (thick lines) with frequency difference 
$(\omega_{z} + \delta)/2\pi$ ($\omega_z$ is Zeeman energy). State $|F = 1, m_{F} = +1 \rangle$ can be neglected when far detuned. 
(b) A representative nonadiabatic dynamic process: The BEC (large red circles) is initially prepared in state $|+\rangle$ with finite momentum and then passes through the avoided crossing point ($k = 0$) and eventually splits into two parts (large red and small blue circles) due to nonadiabatic tunneling. The red dashed and blue solid lines denote the energy spectrum for
states $| + \rangle$ and $|-\rangle$, respectively. Here we use parameters $\Omega/E_r = 0.5$ and $\delta = 0$.} 
\label{model}
\end{figure}

We consider the quantum tunneling problem of an interacting condensate with a SO coupling of equal Rashba and Dresselhaus types \cite{olso, lin1}, described by the Hamiltonian, $\hat{H} = \hat{H}_{\rm SO} + \hat{H}_{\rm trap} + \hat{H}_{\rm int}$. The SO-coupling term is (we set $\hbar \equiv 1$) 
%-----------
\begin{equation} \label{eq101}
   \hat{H}_{\rm SO} = \sum_{\bf k} \begin{bmatrix} \hat{\Psi}_{+}^{\dag}(\bfk) & \hat{\Psi}_{-}^{\dag}(\bfk) 
\end{bmatrix} H_{0}  \begin{bmatrix} \hat{\Psi}_{+}(\bfk) \\
\hat{\Psi}_{-}(\bfk) \end{bmatrix},
\end{equation}
%------ 
with $H_{0} =  \frac{\left(\mathbf{k}  -  {\bf k}_{r} \sigma_z \right)^{2}}{2m} + \frac{\delta}{2} \sigma_z   + \frac{\Omega}{2}\sigma_x$; the trapping potential term is   
%------ 
\begin{equation} \label{eq102}
\hat{H}_{\mathrm{trap}} = - \frac{m \omega^{2}}{2}\sum_{\mathbf{k}, \sigma} \hat{\Psi}_{\sigma}^{\dag} (\mathbf{k}) \nabla_\mathbf{k}^{2} \hat{\Psi}_{\sigma} (\mathbf{k}),
\end{equation}
%-----------
and the interacting term is 
%-----------
\bea \label{eq103}
&&\hat{H}_{\mathrm{int}}  =  \frac{g_\perp}{V} \sum_{\mathbf{k}, \mathbf{k'}, \mathbf{q}} \hat{\Psi}_{+}^{\dag}(\bfk) \hat{\Psi}_{-}^{\dag}(\bfk')\hat{\Psi}_{-}(\bfk'-\bfq)
\hat{\Psi}_{+}(\bfk+\bfq) \notag\\
&&~+ \frac{g_\|}{2V}\sum_{\mathbf{k}, \mathbf{k'}, \mathbf{q}, \sigma} \hat{\Psi}_{\sigma}^{\dag}(\bfk) \hat{\Psi}_{\sigma}^{\dag}(\bfk')\hat{\Psi}_{\sigma}(\bfk'-\bfq)
\hat{\Psi}_{\sigma}(\bfk+\bfq) \notag.
\ea
%-------
Here $\sigma_{x, y, z}$ are the spin-$1/2$ representation of Pauli matrix, and $\sigma = \pm $ denote the two spin states, which are coupled by external fields of an effective Raman coupling strength $\Omega$ and an effective detuning $\delta$. 
${\bf k}_{r}$ is the single-photon recoil momentum, and $\omega$ is the trapping frequency of the harmonic potential. $g_{\|}$ is the $s$-wave interaction strength between the same spin states \cite{note1}, while $g_\perp$ is the interaction strength between the two spins. 
The system Hamiltonian described above has been experimentally realized recently (e.g., see \cite{olso, lin1}), and the associated level diagram is shown in Fig.\,\ref{model}(a). 

Without the trapping potential and the interaction, the single-particle Hamiltonian $\hat{H}_{\rm SO}$ can be easily diagonalized and displays a two-band structure, where the Raman coupling opens a finite band gap, as shown in Fig.\,\ref{model}(b). The typical LZ dynamical problem we will consider is illustrated in Fig.\,\ref{model}(b). The initial condensate wave function is prepared in the $|+\rangle$ state. The harmonic trap potential $\hat{H}_{\rm trap}$ can drive the condensate to move toward the anticrossing point and can lead to a change in the internal spin state due to SO coupling. If the dynamics is adiabatic, all particles shall stay in the upper band and are changed into the $|-\rangle$ state. However, due to the nonadiabatic tunneling between the two energy eigenstates, a certain number of particles are linked to the lower energy level. Furthermore, the condensate wave function may also be affected during the process, as we will show further below.

%========      
%========= 
%%%%%%%%%%%%%%%%%%%%%%%%%%%%%%%%%%%%%%%%%%%%%%%%%%%%%%%%%%%%%%%%%%%%%%  
\section{Quantum limit: multichannel interference} 
\label{quantum_limit}

When the condensate is driven by a weak force, the kinetic energy is much smaller than the oscillating frequency between the different spin states. In such a quantum limit, analytic results may be obtained in the noninteracting case. This can provide a good starting point for further investigation in a more general parameter regime. To get insight and make a comparison with the complex results in the latter part, we begin by exploring the quantum dynamics in the noninteracting SO coupling system, i.e., $g_{\|} = g_{\perp} = 0$. 
%Even in such simple case, the nonadiabatic decay of the condensate can be still caused by two distinct mechanisms: the multichannel quantum interference due to the spin-orbital coupling and the temporal variation of the quantum number (here it refers to the momentum). We will show that the former leads to a special universal power-law decay in the long time limit, while the later can be classified into the celebrated Landau-Zener mechanism in the semiclassical limit. 
For simplicity, we consider one-dimensional(1D) case with zero detuning ($\delta = 0$), although our results should be qualitatively applicable in higher-dimensional systems or with finite detuning.
      
\subsection{Wave function in momentum space}

In the noninteracting limit and when the external harmonic trapping potential is negligible compared to the SO coupling, the energies of the SO-coupling system subject to the Hamiltonian (\ref{eq101}) read 
\begin{equation} \label{eq104}
   \Lambda_{\pm}(k) = E_k + E_r \pm \frac{1}{2}\Omega_k,
\end{equation}
where $\Omega_{k}\equiv\sqrt{16E_k E_r+\Omega^2}$ with the kinetic energy $E_k\equiv k^2/2m$ and the recoil energy $E_r\equiv k_r^2/2m$. To obtain the system wave function at zero temperature, we define $|\Psi(t)\rangle = \sum_{k, \sigma} \Psi_{\sigma}(k,t)|\sigma, k\rangle$ and derive the equation of motion by $\frac{\delta \langle \Psi(t)| \hat{H}_{\rm so} - i \partial_{t} |\Psi(t) \rangle} {\delta \Psi_{\sigma}^{*}(k, t)} = 0$. The final wavefunction $|\Psi(t) \rangle$ can be obtained by calculating $\Psi_{\sigma} (k, t)$. In principle, a similar calculation for such a system can be performed at finite temperature, but we will concentrate on the zero-temperature case here.

After some straightforward algebra assuming all particles are initially prepared in $|+\rangle$ state [i.e., $\Psi_-(k,0)=0$], we have
%--------
\begin{eqnarray}
\Psi_+(k,t)&=&\left[\frac{e^{-i\Lambda_+(k)t}}{1+e^{-2\theta_k}}+\frac{e^{-i\Lambda_-(k)t}}{1+e^{2\theta_k}}\right]\Psi_+(k,0),
\label{Psi+}
\\
\Psi_-(k,t)&=&\frac{{\rm sech}\theta_k}{2}\left[e^{-i\Lambda_+(k)t}-e^{-i\Lambda_-(k)t}\right]\Psi_+(k,0),
\label{Psi-}
\end{eqnarray}
%-------
where we define $\cosh\theta_k \equiv \sqrt{1+16E_k E_r/\Omega^2} = \Omega_k/\Omega$.
Thus, the probabilities of finding a particle remaining in the same spin state ($|+\rangle$) and in the other state ($|-\rangle$) are given by, respectively,  
%----
\bea
   P_+(t) & = & \frac{1}{N} \sum_{k} \left[1-\frac{\Omega^{2}}{\Omega_{k}^{2}} \mathrm{sin}^{2} (\Omega_{k} t/2)\right] |\Psi_{+}(k, 0)|^{2},
         \label{P_eq1}
\\
P_-(t)&=&\frac{1}{N} \sum_{k} \frac{\Omega^{2}}{\Omega_{k}^{2}} \mathrm{sin}^{2} (\Omega_{k} t/2) |\Psi_{+}(k, 0)|^{2} \label{P_eq2},
\ea
%----
where $N = \sum_{k} |\Psi_+(k,0)|^2$ is the total number of particles.

%========              
\begin{figure}[tbp]
\centering
\includegraphics[scale = 0.36]{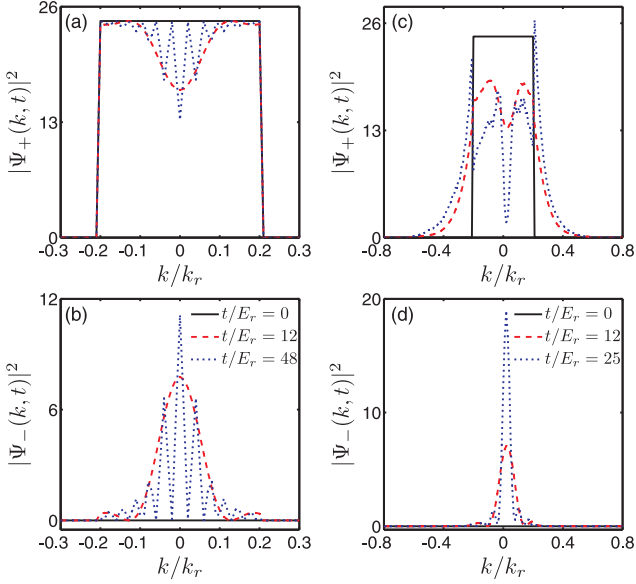} 		        	     
\caption{(Color online) Temporal evolution of momentum density distributions for $|+\rangle$ and $|-\rangle$ components with an initial condition of a uniform distribution ($k_{F} = 0.2k_{r}$). (a) and (b) Noninteracting system;  (c) and (d) the same calculation with $g_{\|} = g_{\perp} = 0.0002E_{r}/k_{r}$. Here $\Omega/E_{r} = 0.1$ and $N = 1000$ for both calculations.}
\label{momentum}
\end{figure}
%========

Equations\,(\ref{P_eq1}) and (\ref{P_eq2}) show that the noninteracting condensate wave functions at different momenta $k$ have different oscillation frequencies as well as oscillation amplitudes. As a typical example, the time evolution of the momentum distributions $|\Psi_{+}(k,t)|^2$ and $|\Psi_{-}(k,t)|^2$ for $\Omega = 0.1 E_r$ are shown in Figs.\,\ref{momentum}(a) and \ref{momentum}(b). Without loss of generality, we consider the initial wavefunction with uniform distribution in the limited momentum space, i.e., $\Psi_{+}(k,0) = \sqrt{N/2k_F}$ for $|k|\leq k_F$ and $0$ otherwise. Figure\,\ref{momentum}(a) shows that in the short-time regime, $|\Psi_{+}(k,t)|^2$ has a sharp valley around $k = 0$ because the atoms in the $|+\rangle$ state are coherently transferred to the $|-\rangle$ state and the tunneling amplitude varies with $1/\Omega_{k}^{2}$. For a sufficiently long time \cite{note2}, a Fresnel interference pattern appears in both bands owing to the momentum-dependent Rabi oscillation term $\sin^2(\Omega_k t)$, giving the width of the central peak as $\Delta k/k_r \sim \sqrt{\pi\Omega/8 t}/E_r$ [by setting $(\Omega_{k = \Delta k}-\Omega_{k=0})t \sim \pi$]. Equations\,(\ref{P_eq1}) and Eq.(\ref{P_eq2}) show that the momentum-dependent oscillation frequency $\Omega_k$ is proportional to $|k|$ in the large momentum regime. As seen explicitly, with increasing $t$, an increasing number of atoms in a relevant small-momentum regime enter the fast oscillation.  Therefore, we can expect the temporal spatial distribution of the condensate wave function may have significant changes due to the fast nonuniform oscillating phase. This will be discussed in the following part.

%%%%%%
\subsection{Dynamical fragmentation of condensate wave function}

%========              
\begin{figure}[bp]
\centering
\includegraphics[scale = 0.4]{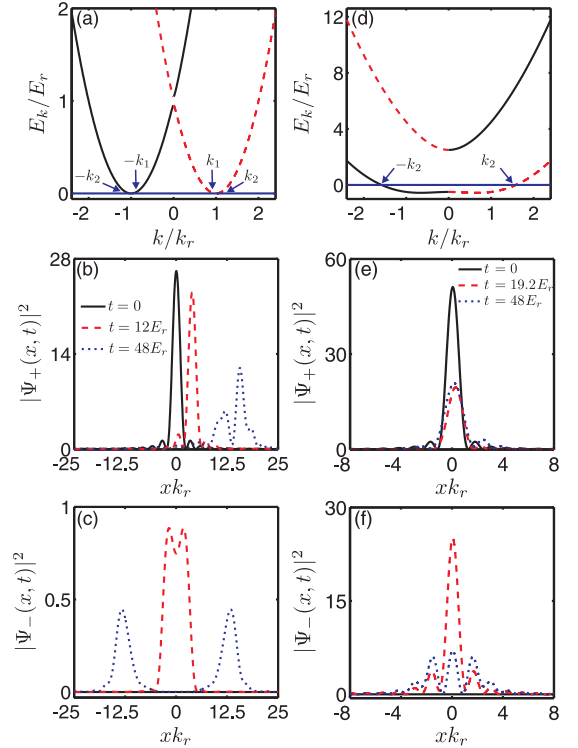} 		        	     
\caption{ (Color online) (a) and (d) The single-particle energy spectrum of the SO-coupled BEC for $\Omega / E_{r} = 0.1$ and $3$, respectively, and the associated time evolution of the density distribution for (b) and (e) $|\Psi_{+}(x, t)|^{2}$ and (c) and (f) $|\Psi_{-} (x, t)|^{2}$. The initial states are the same for both cases with a uniform distribution ($k_{F} = 0.2 k_{r}$), and the other identical parameters are $g_{\|} = g_{\perp} = 0$ and $N = 1000$.} 
\label{real_space}
\end{figure}
%========

Besides the distribution in momentum space, the temporal distribution in real space also manifests interesting properties of multichannel interference. Precisely, the wave function in the $|-\rangle$ state by means of the Fourier transform of Eq.\,(\ref{Psi-}) yields 
%-------
\begin{eqnarray} \label{Psi-real}
\Psi_{-}(x,t) & = & \int \frac{dk}{2\pi}\,e^{ikx}\frac{{\rm sech}\theta_k}{2}
\nonumber\\
&&\times\left[e^{-i\Lambda_+(k)t}-e^{-i\Lambda_-(k)t}\right]\Psi_+(k,0).
\end{eqnarray}
%--------
It is stressed that ${\rm sech}\theta_k = \Omega/\Omega_k$ is peaked at $k = 0$ with a width $\Delta k$ where $16E_{\Delta k} E_r/\Omega^2\sim 1$ and thus $\Delta k\sim m\Omega/2k_r$. The integral (\ref{Psi-real}) is dominantly contributed by the small-$k$ regime, i.e., $|k| < \Delta k$. On the other hand, in the long-time limit, $t\to\infty$, the time-dependent phase term $e^{-i\Lambda_{\pm}(k) t}$ performs fast oscillation if $\Lambda_{\pm}(k) \neq 0$. Hence the major contribution of the integral arises from $\Lambda_{-}(k) \sim 0$; in other words, $k = \pm k_1,\pm k_2$, where $k_{1,2}\equiv k_r\sqrt{1\mp\Omega/2E_r}$ according to the zeros of $\Lambda_-(k)$. [Note $\Lambda_{+}(k) \geq \Omega/2$ is the upper band; see Fig.\,\ref{real_space}(a) and \ref{real_space}(d).]
On the basis of two analysies above, we find that in the weak-SO-coupling regime, $\Omega < 2E_{r}$, $\Psi_{-}(x,t)$ is mostly contributed from $k \sim \pm k_1 = \pm k_r\sqrt{1-\Omega/2E_r}$, and the condensate prefers to be separated by moving in two opposite directions with velocities $\pm v_1$, which is the dispersion velocity near $\pm k_1$ [see Fig.\,\ref{real_space}(c)]. For strong SO coupling, $\Omega > 2E_r$, $\pm k_1$ disappear and $\pm k_2$ are, in general, much larger than the initial momentum distribution of the condensate [see Fig.\,\ref{real_space}(d)]. 
While all the phases in each momentum channel oscillate quickly, the distinction of oscillation frequencies in different momentum channels is greatly suppressed. Therefore, the real-space condensate wave stops moving away, and its amplitude varies quickly, but the configuration remains highly in contrast to weak SO coupling for a relatively long time [for example, for $\Omega = 3$ the shape of the condensate distribution holds well up to $t =  20 E_{r}$; see Fig.\,\ref{real_space}(f)].

For weak SO coupling, the situation for the $|+\rangle$ state is distinct from that of the $|-\rangle$ state, as shown in Fig.\,\ref{real_space}(b): for $\Omega=0.1 E_r$ the condensate of the $|+\rangle$ state moves in the positive $x$ direction and gradually becomes fragmented as time goes on. This can be easily interpreted from the expression 
%-------
\begin{align*}
\Psi_{+}(x,t)& = \int \frac{dk}{2\pi}\,e^{ikx}\left[\frac{e^{-i\Lambda_+(k)t}}{1+e^{-2\theta_k}}+\frac{e^{-i\Lambda_-(k)t}}{1+e^{2\theta_k}}\right]\Psi_+(k,0).
\end{align*}
%--------
Again, the dominant contribution comes from the momentum near $\pm k_{1,2}$ according to the oscillation of the phase factor $\Lambda_{-}(k) t$. In contrast to Eq.\,(\ref{Psi-real}), the weighting function from the dominant term behaves differently: the second term above is proportional to $(1+e^{2\theta_k})^{-1}$, which contributes remarkably only in the negative-$k$ regime. As a result, most of the contribution of the condensate wave function in the long-time limit is from $k\sim -k_1$, which gives a positive velocity of the condensate $v_1$. Since the momentum-dependent frequencies in the multichannel tunneling become more apparent for weak SO coupling, there are still finite contributions from other momentum channels, which makes the condensate start to fragment with different velocities in different spaces. 
However, the momentum-dependent frequencies are suppressed greatly for strong SO coupling, and correspondingly, the momentum values which fulfill $\Lambda_{-}(k) = 0$ are larger than $k_{F}$, e.g., $\Omega > 2E_{r}$ for $k_{F} = 0.2k_{r}$, so that the dynamic behavior for $|\Psi_{+}(x, t)|^{2}$ is similar to that for $|\Psi_{-}(x, t)|^{2}$ for a finitely long time; for example, for $\Omega = 3$, the similar dynamic behavior holds well up to $t = 20 E_{r}$ [see Fig.\,\ref{real_space}(e)].

%%%%%%%%%%%%%
\subsection{Universal survival probability \texorpdfstring{$P_+(t)$}{P(t)}}

%========
\begin{figure}[bp]
\centering
\includegraphics[scale = 0.6]{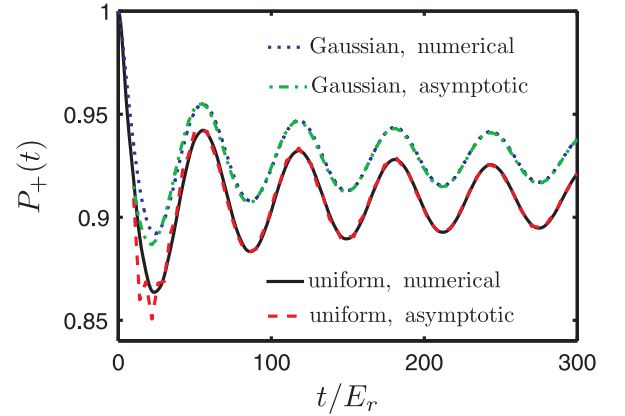} 		        	
\caption{(Color online) Comparison of numerics and the analytic formula (\ref{Pt_analytic}). The numerical calculation of the survival probability $P_+(t)$ for uniform distribution (black solid line) and Gaussian distribution (blue dotted line) of a noninteracting system in the initial wave function in momentum space. The initial uniform distribution has $k_{F} = 0.2 k_{r}$, and the Gaussian distribution, $\sim e^{-k^{2}/2\kappa^{2}}$, has a width $\kappa = 0.2k_{r}$. The analytic asymptotic formulas of (\ref{Pt_analytic}) corresponding to the two cases are displayed with red dashed and green dash-dotted lines, respectively. The value of $\alpha$ for the latter is obtained by fitting the long-time behavior and is equal to $0.39$ in this plot. The other identical parameters are $\Omega/E_{r} = 0.1$ and $N = 1000$. 
}
\label{Fig9}
\end{figure}
%========

It is preferable and relatively easy for experimentalists to measure the nonadiabatic dynamics of SO-coupled BEC by means of counting the number of particles in the different spin states. Since $P_-(t) = 1 - P_+(t)$, we merely investigate the analytic expression of the survival probability $P_+(t)$ in both short-time and long-time limits.

As $t\to 0$, the asymptotic solution of Eq.\,(\ref{P_eq1}) shows $P_{+}(t)=1 - \Omega^2t^2/4+{\cal O}(t^4)$, which matches a typical Rabi oscillation. Note that in our system, the traditional single-channel Rabi oscillation frequency is $\Omega$, while the multichannel Rabi frequency in the short-time behavior is $\Omega/\sqrt{2}$. In the relatively long-time limit, for simplicity, we just consider the generic case, where the initial density distribution in momentum space is a constant up to $|k| < k_F$, as mentioned above. We point out that a more complicated initial wave function merely quantitatively affects the asymptotic behavior and can be easily adjusted by a single fitting parameter, as shown below. Within the uniform initial wave function, we can evaluate the summation in the momentum channel by integration and derive the leading-order expression in the long-time limit by considering only the major contribution. After some nontrivial calculations (see the Appendix for details), we find that
%----
\bea
   P_{+}(t) & = & 1 - \frac{\alpha}{8}\tan^{-1}\left(\frac{4}{\alpha}\right)
              + \frac{\sqrt{2\pi}\alpha}{16}\frac{\cos(\Omega t+\pi/4)}{\sqrt{\Omega t}}
\nn
            &   & + \frac{\alpha^2}{32}\frac{\sin(\Omega_{k_F}t)}{\Omega t}+O(t^{-3/2}),
\label{Pt_analytic}
\ea
%----
where $\alpha\equiv\Omega/\sqrt{E_FE_r}$ is a dimensionless parameter and $E_F\equiv E_{k_F}$ is the ``effective Fermi energy.''

There are several interesting observations from Eq.\,(\ref{Pt_analytic}): (i) The multichannel quantum interference effects can also lead to a ``decay" with an oscillating amplitude, which is a special feature of the SO-coupled condensate due to the momentum dependence of the Rabi oscillation frequency $\Omega_k$. (ii) The saturated value $P_{+}(\infty) = 1-\frac{\alpha}{8}\tan^{-1}\left(\frac{4}{\alpha}\right)$ approaches $1/2$, as expected for a single-mode Rabi oscillation if $E_F\ll \Omega$ ($\alpha\to 0$). On the other hand, it reaches $P_{+}(0) = 1$ if $E_F\gg\Omega$ ($\alpha\to 0$), thus indicating that the energy band width (uncertainty in energy) can reduce the many-body quantum tunneling through interference effects. (iii) These results can also be applied to noninteracting fermions since the Pauli exclusion principle requires $\Psi_+(k,0)=1$ for $|k|\leq k_F$ and the sign of wave function exchange should not affect the probability of finding a particle in any spin channel. (iv) For a general initial wave function, $\Psi_+(k,t)$, the long-time behavior of $P_{+}(t)$ can be still applied, except one can use $\alpha$ as a fitting parameter, characterizing the energy uncertainty of the initial wave function in the momentum space.
In Fig.\,\ref{Fig9}, we show the time dependence of the survival probability $P_{+}(t)$ as a function of time for both uniform distribution and a Gaussian distribution in the initial wave function. Results from the analytic expression [Eq.\,(\ref{Pt_analytic})] are also shown together (the value of $\alpha$ for the Gaussian distribution is given by single-parameter fitting). The analytic results for both distribution agree with the corresponding numerical results as $t>\Omega^{-1}$.
 
%%%%%%%%%%%  
\section{Classical limit: Landau-Zener tunneling}
\label{classical_limit}

Now we consider another region in which the ``kinetic energy'' arising from the external trapping potential dominates the nonadiabatic dynamics. Here we will employ a semiclassical approach in this region and show its connection with the well-known Landau-Zener effect. In the semiclassical limit, one can treat the center-of-mass position, $x(t)$, and momentum, $k(t)$, as a classical particle at time $t$ and neglect the spatial or momentum distribution induced by the condensate wave function. As a result, the system dynamics can be described by a two-component state $[\psi_+(t),\psi_-(t)]^T$, which is controlled by the Hamiltonian in Eq.\,(\ref{eq101}) in the moving frame of momentum $k(t)$:
%-----
\begin{equation}
i \partial_{t} \begin{bmatrix} \psi_{+} \\ \psi_{-} \end{bmatrix} = \begin{bmatrix} \frac{(k(t)-k_r)^2}{2m} & \Omega/2 \\
\Omega/2  & \frac{(k(t)+k_r)^2}{2m} \end{bmatrix}
            \begin{bmatrix} \psi_{+} \\ \psi_{-} \end{bmatrix},
\label{LZ_semiclassical}
\end{equation}
%-----
where the ``quasimomentum'' $k(t)$ becomes a time-dependent external parameter, controlling the time dependence of the single-particle energies as a typical LZ-type problem. If the condensate is initially prepared in momentum state $k(0)=-k_0<0$ of the $|+\rangle$ state [see Fig.\,\ref{model}(b)], the time dependence of $k(t)$ in the semiclassical approximation fulfills $k(t) = -(k_0+k_r)\cos(\omega t)+k_r$ in a harmonic potential well.

Following the standard treatment of LZ tunneling, we perform the time variation of $k(t)$ approximately as a linear function of $t$ when the condensate is at the $k=0$ ($t=t_c$) point. From $k(t_c)=0$ we can have $\cos(\omega t_c)=k_r/(k_r+k_0)$. The probability for such a particle to be transferred into the $|-\rangle$ state after a long time has the form,
$P_{\mathrm{LZ}}(\infty) = \mathrm{exp} \left[ - \frac{2\pi \left(\Omega/2\right)^{2}} {\left|(dE_+(k) - dE_{-}(k))/dt\right|_{k =0}}\right]$, where $E_\pm(k) = \frac{(k\mp k_r)^2}{2m}$ and thus $\left|\frac{d\left[E_+(k) - E_{-}(k)\right]}{dt} \right|_{k=0} = \frac{2k_r}{m}\left|\frac{dk}{dt}\right|_{t=t_c}$. Since
$\left|\frac{dk}{dt}\right|_{t=t_c} = \omega\sqrt{(k_r+k_0)^2-k_r^2}$, we have
%----
\begin{eqnarray}\label{eq3}
P_{\mathrm{LZ}}(\infty) = \mathrm{exp} \left[- \frac{\pi\Omega^{2}}{8\omega E_{r}\sqrt{(k_0/k_r)^2+2(k_0/k_r)}}\right].
\end{eqnarray}
%-----
In contrast to quantum limit in which the diabatic state undergoes an oscillating power-law decay in terms of $\Omega$ due to multichannel interference, [see Eq.\,(\ref{Pt_analytic})], in the semiclassical treatment, the formula (\ref{eq3}) indicates that when such state is driven to pass the anticrossing of energy bands, it will experience the exponential decay with respect to $\Omega$.

%========
\begin{figure}[htbp]
	 \centering
	 \includegraphics[scale = 0.5]{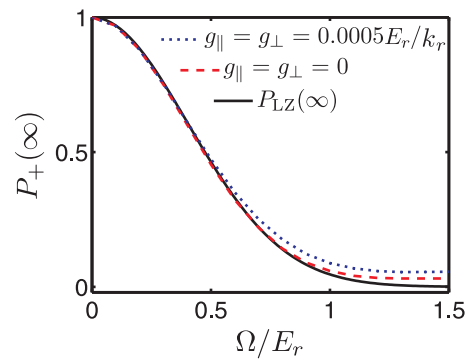} 		        	     
	 \caption{(Color online) Comparison of numerics and semiclassical analysis. The result evaluated for the classical Landau-Zener formula, 
[see Eq.\,(\ref{eq3})], according to our system parameters (black solid line) is compared with full quantum mechanics calculations for noninteracting  (red dashed line) and interacting (blue dotted line) SO-coupled BECs. The initial nonequilibrium state is chosen to be $\Psi_{+}(k, 0) \sim  \rm{exp}\left[- (k + k_{0})^{2}/(4 \kappa^{2}) \right]$, where $k_{0} = k_{r}$ and $\kappa = 0.2 k_{r}$. The other identical parameters are $\omega = 0.08E_{r}/\hbar$ and $N = 1000$.} 
	 \label{LZ}
\end{figure}   
%========
Figure\,\ref{LZ} shows typical results of $P_{+}(\infty)$ from numerically solving Eq.\,(\ref{eq2}) and comparing with the semiclassical analysis derived above. Our results show that the LZ formula matches the full numerical results very well in the weak-coupling regime, while it deviates in the large $\Omega$ limit. 
This might be due to the following reason. The large $\Omega$ broadens the effective momentum regime where the atoms experience multichannel interference [see Eqs.\,(\ref{P_eq1}) and (\ref{P_eq2})]. As a consequence, the multichannel tunneling is involved in the LZ tunneling procedure, and the assumption of a linear time dependence of $k(t)$ at the anticrossing region becomes inaccurate in such a limit.

%%%%%%%%%%%%%%%%%%%%%%%%%%%%%%%%%%%%%%%%%%%%%%%%%%%%%%%%%%
\section{Interacting spin-orbit-coupled BEC}  \label{secIV}      

Now we explore the interaction effects on the nonadiabatic dynamics. Following the standard approach assuming all particles condensate in a single-particle wave function, i.e., $\langle\hat{\Psi}_\sigma\rangle=\Psi_\sigma$, the mean-field Gross-Pitaevskii equation (GPE) can be written as follows:
%------
\begin{align}\label{eq2}
i \partial_{t}  \Psi_{\pm}& (k)  = \left[ \frac{(k \mp k_r)^{2}}{2m}-\frac{m\omega^2}{2}\frac{\partial^2}{\partial k^2} \right] \Psi_{\pm} (k) + \frac{\Omega}{2} \Psi_{\mp} (k) \notag\\
& + \frac{g_\perp}{L} \sum_{k', q} \Psi_{\mp}^{*} (k') \Psi_{\mp}(k'-q) \Psi_{\pm} (k+q) \notag\\
& + \frac{g_\|}{L} \sum_{k', q} \Psi_{\pm}^{*} (k')\Psi_{\pm}(k'-q)\Psi_{\pm} (k+q),
\end{align}
%-----------
where $\sum_{k, \sigma} |\Psi_\sigma(k)|^2 = N $.

There are two temporal effects raised by the contact interaction : the nonlinear dynamics induced by the variation of the mean-field energy and the rearrangement of the particle density distribution due to the scattering between different momentum channels. In what follows, we will show how the contact interaction acts on multichannel interference and LZ tunneling.      
%%%%%%%%
\subsection{Nonlinear dynamics in the single-channel Rabi oscillation}

%=========                            
\begin{figure}[tbp]
\centering
\includegraphics[scale = 0.5]{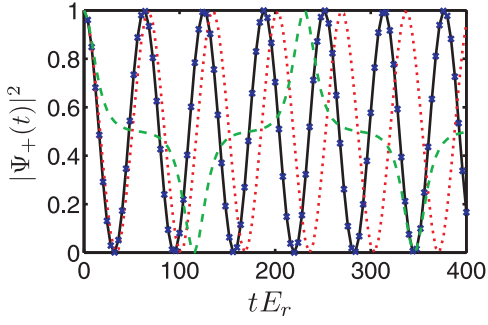} 		        	     
\caption{(Color online) Raman-assisted tunneling dynamics, obtained from Eq.\,(\ref{eq_8}). 
The initial state is $\Psi_{+}(t) = 1$ and $\Psi_{-}(t) = 0$ with $\Omega = 0.1$ for all curves. $g_\| = g_\perp = 0$ for the black solid line, $g_\| =2 g_\perp = 0.2$ for the red dotted line, $g_\| = g_\perp = 0.2$ for the blue crosses, and $g_\| = 0.5 g_\perp = 0.2$ for the green dashed line. 
}
\label{Fig4}
\end{figure}            
%=========

Although our paper concentrates on the multichannel effects in momentum space, it is still instructive to compare our results with those of single-channel (single-mode) dynamics, especially when the interaction is included. The most relevant example assumes all particles condensate at the $k=0$ state in momentum space without an external trapping potential. The corresponding mean-field dynamical equation can be written as follows:
%-------
\begin{equation} \label{eq_8}
  i \partial_{t} \Psi_{\pm} = \left[ g_\| |\Psi_{\pm}|^{2} +{g_\perp} |\Psi_{\mp}|^{2}
  \right] \Psi_{\pm}  + \frac{\Omega}{2} \Psi_{\mp},
\end{equation}
%-------
which is nothing but coupled nonlinear equations, where we have removed the constant term $E_r \Psi_{\pm}$. Here $\Psi_\pm(t)\equiv\Psi_\pm(k=0,t)$ is the wave function at zero momentum. Since the kinetic energy completely disappears, the SO effects turns off, and the Raman coupling term makes the system equivalent to a double-well system with finite tunneling. Both semianalytic and numerical results have been developed in such a context \cite{smer}.

Figure\,\ref{Fig4} shows typical dynamics as a function of time for various strengths of interaction. It can be seen readily that the interaction remains dynamically ineffective when the ${\rm SU}(2)$ symmetry is preserved, i.e., $g_\perp = g_\|$. When the symmetry is not preserved, as in most cases with pseudospins, the tunneling dynamics changes in various ways, depending on the relative strength of the two interactions. In contrast to multichannel tunneling, however, all of these dynamics are coherent and periodic because no decoherence or dissipation is involved in such two-component systems. Such a simple two-component model can be extended to the case of a few components. Interaction merely plays the role of nonlinearity, which is not the case for multichannel systems, as shown in a SO-coupled condensate.

%%%%%%%%%%
\subsection{Multichannel scattering in the quantum interference limit}

%=========                            
\begin{figure}[tbp]
\centering
\includegraphics[scale = 0.5]{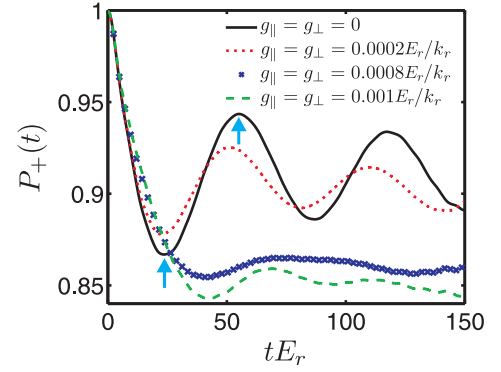} 		        	     
\caption{(Color online) Illustration of the influence of contact interaction on multichannel tunneling by means of measuring the survival probability $P_{+}(t)$ for the uniform distribution of initial wave function in momentum distribution ($k_{F} = 0.2k_{r}$).  $g_\|= g_\perp = 0$ for the black solid line, $g_\| = g_\perp  = 0.0002/k_{r}$ for the red dotted line, $g_\| = g_\perp  = 0.0008/k_{r}$ for the blue crosses, and $g_\|  =  g_\perp  = 0.001/k_{r}$ for the green dashed line. The other parameters are identical to those in Fig.\,\ref{Fig9}. The points at time $t = \pi/\Omega$ and $2\pi/\Omega$ are indicated by the lower left and upper right arrows, respectively.  
}
\label{multinter}
\end{figure}            
%=========

In addition to the mean-field-type energy shift described above, interparticle interaction can also scatter particles in different momentum channels, which makes the multichannel dynamics much more complicated than the noninteracting case. To get insight, we begin by exploring the interaction effect on the condensate dynamics without a trapping potential (i.e., no driven force).

Figures\,\ref{momentum} (c) and \ref{momentum}(d) display the temporal momentum distribution with a finite interaction strength. In the short-time limit, the repulsive interaction certainly broadens the initial uniform distribution to momentum larger than $k_F$, so that the initial distribution of the jump type around $k_{F}$ becomes smooth [see Fig.\,\ref{momentum}(c)]. At late times, the competition between more pronounced tunneling occurring around $k = 0$ [see the amplitude term of Eq.\,(\ref{P_eq1})] and the atoms scattered preferably to the low-density region lead to the disappearance of the Fresnel oscillation pattern shown in Figs.\,\ref{momentum}(a) and \ref{momentum}(b).

In Fig.\,\ref{multinter}, we show the survival probability $P_+(t)$ with respect to various interaction strengths. Compared to the oscillating power-law decay in the noninteracting case (see Fig.\,\ref{Fig9}), an additional time scale for the decay of $P_{+}(t)$, as well as arriving at the dynamic equilibrium state, is required. Specifically, in the initial ``collapse'' of $P_{+}(t)$, i.e., $t < \pi/\Omega$, the repulsive interaction scatters the atoms into high-momentum channels where the tunneling amplitude is approximately reduced by $1/k^{2}$ [see Eq.(\ref{P_eq1})], and thereby it suppresses the whole tunneling. This is manifested in Fig.\,\ref{multinter}, where for $t < \pi/\Omega$, $P_{+}(t)$ for the interacting case is conventionally larger than for the noninteracting case. Afterwards, due to SO coupling, the greater number of atoms tunneling around the low-momentum regime than around the high-momentum regime results in a low-density region around $k = 0$ [see the valley of $|\Psi_{+}(k, t)|^{2}$ shown in Fig.\,\ref{momentum}(c)], while the interaction prefers to scatter atoms into the low-density region. Consequently, the repulsive interaction enhances the whole tunneling of the system, and $P_{+}(t)$ for the interacting case is always smaller than for the noninteracting case for $t > 2\pi/\Omega$, as shown in Fig.\,\ref{multinter}.     
% leading to an exponentially decay ($\sim e^{-t/\tau}$) in the long time limit. The new time-scale, $\tau$, is certainly inverse proportional to the interaction energy, showing the third mechanism to cause non-adiabatic dynamics through multi-channel scattering in SO coupled BEC.

%%%%%%%%%%
\subsection{Interaction effect in the classical limit}

%=========
\begin{figure}[bp]
\centering
\includegraphics[scale = 0.39]{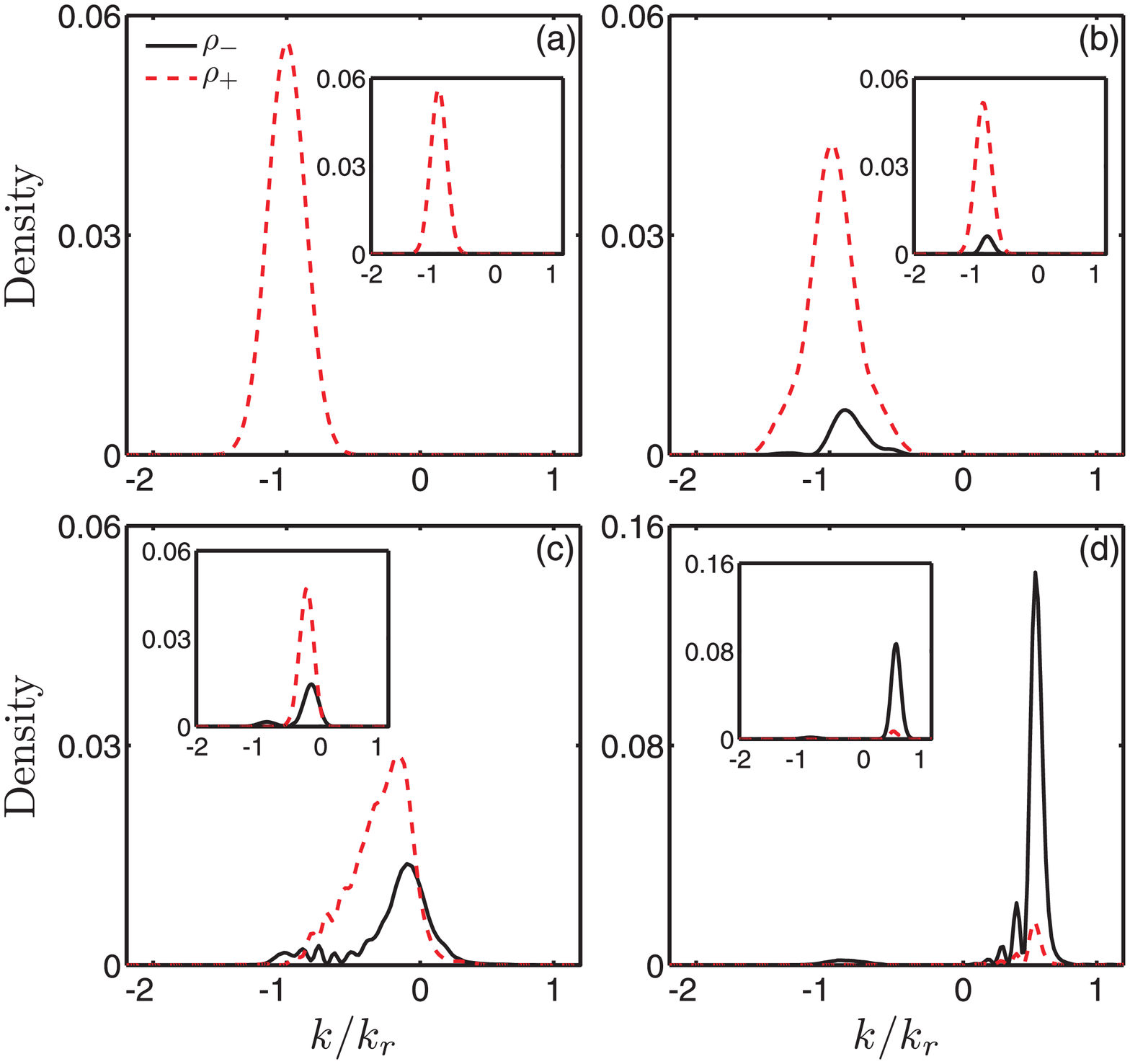} 		        	
\caption{(Color online) Temporal momentum density distribution of a condensate driven by external potential with an initial momentum, $k_0 = -k_r$. 
Red dashed and black solid lines display the density distribution for the $|+\rangle$ and $|-\rangle$ states, respectively,  at time $tE_r$ eual to (a) $0$,  (b) $2.4$, (c) $12$, and (d) $19.6$. The associated parameters are $\Omega/E_r = 1.6$, $g_{\|} = g_{\perp} = 0.001 E_{r}/k_{r}$, and $N = 1000$. For comparison, the insets are the momentum distribution at the same time under identical parameters but without interaction.}

\label{LZint}
\end{figure}
%=========

In this part, we explore the interaction effects on a general driven SO-coupled BEC. As shown above, there are two fundamentally distinct mechanisms for the nonadiabatic dynamics: the multichannel interference in the quantum limit and the LZ effects in the semiclassical (single-channel) regime. Note that in the latter, the whole condensate is treated as a classical particle with definite position and momentum at the same time, indicating that the quantum many-body effects are neglected. However, in any realistic situation, the condensate always has a finite distribution in momentum space, and hence the interparticle interaction can still scatter particles between momentum channels. This implies that the full tunneling dynamics are, very likely, mixed with the two mechanisms shown above. Since such complicated dynamics cannot be readily studied using an analytic approach, here we show the numerically simulated results by solving the full GPEs within the mean-field approximation.

Figure\,\ref{LZint} shows a typical tunneling dynamics of an interacting driven condensate with SO coupling. As we can see, when the condensate wave function approaches the anticrossing point (i.e., $k=0$), particles start tunneling to the $|-\rangle$ state. For the noninteracting case, the original Gaussian shape in both $|+\rangle$ and $|-\rangle$ channels is basically kept because the noninteracting Hamiltonian [see Eq. (\ref{eq101})] is just like a particle moving a simple harmonic potential in momentum space as $(k\pm k_r)/2m\gg \Omega$ in the long-time limit, making each component of a condensate oscillate with the same frequency. This is true even when part of the condensate is split into the other spin state through SO coupling.

However, in comparison with the noninteracting condensate, the density profile of the condensate with the inclusion of a finitely strong interaction is displayed in a different way as it passes the critical point. When the interaction is increased from zero, the condensate density distribution becomes highly distorted after passing through the critical point: the head of the condensate is compressed to a much narrower peak, while the tail is destroyed without any smooth profile. Such results can be understood from the scattering between multichannel momenta: the condensate is broadened from its initial profile by the repulsive interaction during the motion (note that the noninteracting condensate profile is not broadened in momentum space), so that the ``velocity'' of the wave head is faster than the ``velocity'' of the wave tail, making many more particles tunnel into the $|+\rangle$ state in the tail part. The small density oscillation shows the interference effect of two such ``velocities'' in momentum space. Moreover, since the two-body interaction enhances the multichannel effect, the tunneling rate for the interacting SO BEC deviates more from classical LZ rate than in the noninteracting case (see Fig.\,\ref{LZ}). We propose that the suppressed tunneling rate and the condensate distortion could be a feature of many-body effects and could be observed in current experiment setup \cite{olso, lin1}.

%We further explore the interaction effects of the multi-channel on-adiabatic dynamics of the SOC condensate by including the quantum fluctuations via using Truncated Wigner approximation (TWA) \cite{stee, blak}. We note that the relevant classical process is governed by deterministic time-dependent GP equation, while TWA includes leading order quantum fluctuations by sampling the classical trajectories by Wigner representation, i.e. $\Psi_{\pm}(k, t = 0) = \Psi_{\pm, MF}(k) + \xi_{\pm}(k)$, where $\Psi_{\pm, MF}(k)$ is the initial mean-field wave function and $\xi_{\pm}(k)$ is interpreted as complex Gaussian random field. More details can be found in Ref \cite{???}.

%In Fig. \ref{???} we show the calculation of condensate profile calculated by TWA (as a comparison, we also show noninteracting and MF results together). As one can see, Including quantum fluctuations gives qualitatively the same profile as the meanfield calculation. Such a result indicates that the multi-channel scattering of condensate during the non-adiabatic tunnelling can be also understood as a "decoherence" effect, because wavefunction coherence at different components (assumed in the meanfield approximation) is almost irrelevant. This is very different from the noninteracting multi-channel regime, semi-classical limit, or even interacting regime with single channel only (dicussed above). This confirms our scenario that the full non-adiabatic dynamics has to be understood within the multi-channel quantum interference.

%%%%%%%%%%%%%%%%%%%%%%%%%%%
\section{Experimental parameters and summary} \label{secV}

     In principle, any geometry and dimension of the condensates with SO coupling are applicable to exploring experimentally multichannel tunneling dynamics and interaction-induced collapse. We propose a 1D or quasi-1D configuration, which is probably optimal for experimentalists to observe the phenomena by controlling and adjusting physical parameters because fewer parameters are needed than in higher dimension, in accordance with our calculation. We consider a $^{87}$Rb BEC with the $s$-wave scattering length $a = 5.4$nm, confined in the harmonic potential with the transverse frequency $\omega_{y} = \omega_{z} = 2\pi \times 1000$Hz and the longitudinal frequency $\omega_{x} = 2\pi \times 160$Hz. Supposing that the Raman coupling laser has a wavelength $\lambda = 804.3$nm \cite{lin1, olso}, the quasi-1D effective scattering amplitude is approximately $g_{1D} = 2 \hbar a \sqrt{\omega_{y}\omega_{z}} \approx 0.02 E_{r}/k_{r}$. In our calculation, $g_{\|} = g_{\perp} \sim g_{1D}/2\pi$, so $g_{\|} = g_{\perp} \sim 0.003 E_{r}/k_{r}$. This type of interaction coupling can match our calculation; note that the mean-field dynamics are essentially determined by $g_{\|}N$ and $g_{\perp}N$, and thus the total particle number $N$ can also be adjusted to control the dynamics.     
        
   In summary, we investigate the full dynamics of a BEC subject to SO coupling. We show that SO coupling is a special technique to address and to detune the tunneling rate in each momentum channel. The corresponding complete nonadiabatic dynamics for a SO BEC is thus much richer than the traditional LZ mechanism, even in the absence of contact interaction. The quantum interference between different momentum channels and scattering between them give phenomena, which are qualitatively different from ones from single-channel mechanism. We show how these different mechanisms can be addressed in different experimental regimes within the same frame work. Our prediction on the power-law decay of the transition rate as well as the wave function fragmentation due to multichannel quantum interference can be observed within the current experimental setup.

%%%%%%%%%%%%%%%%%%%%%%%%%%%%%%%%%%%%%%%%%%%%
\acknowledgments
We thank H. Zhai, and Y.-J. Lin for the inspiring discussions. The work was supported by NCTS and MoST in Taiwan. B. X. also acknowledges support from NSFC Grant No. 11347025. 

\appendix

\section{Power-law decay of the survival rate for the condensate with uniform distribution in momentum space} \label{math}
   The survival probability of noninteracting particles in the initial spin state yields
%------
   \begin{equation} \label{app_eq1}
     P(t) = \frac{1}{N} \sum_{{\bf k}} \left[ 1 - \frac{\Omega^{2}}{\Omega_{k}^{2}} \mathrm{sin}^{2} (\Omega_{k} t/2) \right] |\Psi_{+}(k, 0)|^{2},
   \end{equation}
%------
where $\Omega_{k} = \sqrt{16 E_{k} E_{r} + \Omega^{2}}$ with $E_{k} \equiv k^{2}/2m$, and $E_{r} \equiv k_{r}^{2}/2m$. 

Considering $\Psi_{+}(k, 0) = \sqrt{\frac{N}{2k_{F}}}$ for $|k| \leq k_{F}$ and is otherwise zero, Eq.\,(\ref{app_eq1}) can be written as
%------
   \bea \label{app_eq2}
         P(t) & = & \frac{1}{k_{F}} \int_{0}^{k_{F}} dk \left[1 - \frac{\Omega^{2}}{2 \Omega_{k}^{2}} \left( 1 - {\rm cos} (\Omega_{k} t) \right) \right] \nonumber \\
              & = & 1 - \frac{1}{8} \sqrt{ \frac{\Omega^{2}}{E_{F} E_{r}} } {\rm tan}^{-1} \left( \frac{4 \tilde{k}_{F}} {\tilde{\Omega}} \right) + \mathcal{A},
   \ea
%------
   where $\tilde{k} \equiv k/k_{r}$ and $\tilde{\Omega} = \Omega/E_{r}$. The critical oscillation part has the form
   \begin{equation} \label{app_eq2a}
     \mathcal{A} = \frac{1}{\tilde{k}_{F}} \int_{0}^{\tilde{k}_{F}} d\tilde{k} \left( \frac{\tilde{\Omega}^{2}}{2 \tilde{\Omega}_{k}^{2}} \right) {\rm cos} (\tilde{\Omega}_{k} \tilde{t}),
   \end{equation} 
where $\tilde{t} = t E_{r}$ and $\tilde{\Omega}_{k} = \sqrt{16 \tilde{k}^{2} + \tilde{\Omega}^{2}}$.
   
   The following procedure is to obtain the analytic expression for $\mathcal{A}$ in some proper treatment. We arrange $\mathcal{A}$ into
%------   
   \begin{equation} \label{app_eq3}
      \mathcal{A} = \frac{\tilde{\Omega}} {8 \tilde{k}_{F}} \int_{1}^{\sqrt{ 1 + \left( \frac{4 \tilde{k}_{F}} {\tilde{\Omega}} \right)^{2}}} dy 
                    \frac{ {\rm cos}(\tilde{\Omega} \tilde{t} y)} {y \sqrt{y^{2} -1 }},
   \end{equation}
%------   
   where $y = \sqrt{ 1 + \left( 4 \tilde{k}/ \tilde{\Omega} \right)^{2}}$. To denote our calculation in a convenient way, we further define
   \begin{equation} \label{app_eq2b}
      F(\tilde{t}, x) = \int_{x}^{\infty} dy \frac{\mathrm{cos}(\tilde{t} y)} {y \sqrt{y^{2} - 1}} dy
   \end{equation}
   with $\tilde{\Omega} \tilde{t} \rightarrow \tilde{t}$, so
   \begin{equation} \label{app_eq2c}
      \mathcal{A} = \frac{\tilde{\Omega}}{8 \tilde{k}_{F}} \left[ F(\tilde{t}, 1) - F(\tilde{t}, a) \right],
   \end{equation}
   where $a = \sqrt{1 + \left(4\tilde{k}_{F}/\tilde{\Omega} \right)^{2}}$. 
   For $F(\tilde{t}, a)$ with $a > 1$, we can expand the square-root term and obtain
   \bea 
      F(\tilde{t}, a) & = & \int_{a}^{\infty} dy \frac{\mathrm{cos} (\tilde{t} y)} {y^{2}} \left[ 1 + \frac{1}{2} \frac{1}{y^{2}}
                            + \frac{3}{8} \frac{1}{y^{4}} + \cdots \right] \nn
                      & \equiv & F_{2}(\tilde{t}) + \frac{1}{2} F_{4} (\tilde{t}) + \frac{3}{8} F_{6} (\tilde{t}) + \cdots  \label{app_eq2d},
   \ea
where $F_{2n}(\tilde{t}) \equiv \int_{x}^{\infty} \frac{\mathrm{cos}(\tilde{t}y)}{y^{2n}} dy$.  Let $z = \tilde{t}y$, so
   \bea 
     F_{2n} (\tilde{t}) & = & \tilde{t}^{2n - 1} \int_{a \tilde{t}}^{\infty} \frac{\mathrm{cos}z }{z^{2n}} dz  \nn
                        & = & \frac{\tilde{t}^{2n-1}}{2} \left[ \int_{a \tilde{t}}^{\infty} (e^{iz - \eta z} + e^{-iz - \eta z}) z^{-2n} \right] dz, \nonumber
   \ea 
where $\eta \rightarrow 0^{+}$ makes the integral convergent.  

Since $\int_{a \tilde{t}}^{\infty} z^{-2n} e^{- (\eta \mp i) z}  dz = (\pm i)^{1- 2n} (\mp i a \tilde{t})^{-2n} e^{\pm i a \tilde{t}}$ for $\tilde{t} \rightarrow \infty$, 
   \begin{equation} \label{app_2f}
      F_{2n} (\tilde{t}) = - \frac{1}{a^{2n}} \frac{\mathrm{sin} (a \tilde{t})}{\tilde{t}} + O\left( \tilde{t}^{-2} \right).
   \end{equation}
Correspondingly, 
   \begin{equation} \label{app_2g}
      F(\tilde{t}, a) = - \frac{\mathrm{sin} (a \tilde{t})}{\tilde{t}} \frac{1}{\sqrt{a^{2} - 1}} + O\left( \tilde{t}^{-2} \right).
   \end{equation}    
   
   For $F(\tilde{t}, 1)$ and when $\tilde{t} \rightarrow \infty$, the ``high-frequency'' wave according to large $y$ can be approximately neglected, and the part of small $y$, i.e., $y \rightarrow 1$, dominates the integral. So we can treat approximately
%------   
   \begin{equation} \label{app_eq4}
     F(\tilde{t}, 1)  \approx \frac{1}{\sqrt{2}}\int_{1}^{\infty} dy \frac{ {\rm cos}(\tilde{t} y)} {\sqrt{y -1 }}.           
   \end{equation}
%------   
As $\tilde{t} \rightarrow \infty$, Eq.(\ref{app_eq4}) has the solution
%------
   \begin{equation} \label{app_eq5}
      F(\tilde{t}, 1) = \sqrt{\frac{\pi}{\tilde{t}}} \mathrm{cos} \left( \tilde{t} + \frac{\pi}{4} \right) + \mathcal{O} 
      \left(\tilde{t}^{-3/2} \right) 
   \end{equation}  
%------    
Through Eqs.\,(\ref{app_eq2}), (\ref{app_eq2c}), (\ref{app_2g}), and Eq.(\ref{app_eq5}) , one can obtain, at $t \rightarrow \infty$, 
%------
\begin{eqnarray}
P(t) & = & 1-\frac{\alpha}{8}\tan^{-1}\left(\frac{4}{\alpha}\right)
       + \frac{\sqrt{2\pi}\alpha}{16}\frac{\cos(\Omega t+\pi/4)}{\sqrt{\Omega t}} \nonumber\\
     &   & +\frac{\alpha^{2}}{32}\frac{\sin(\Omega_{k_F}t)}{\Omega t}+{\cal O}(t^{-3/2})
     \label{app_eq6}
\end{eqnarray}
%----
where $\alpha\equiv\Omega/\sqrt{E_FE_r}$ is a dimensionless parameter to control the decaying amplitude and $E_F\equiv E_{k_F}$ is the effective Fermi energy.

%%%%%%%%%%%%%%%%%%%%%%%%%%


\begin{thebibliography}{000}
\bibitem{note1} The assumption that the $s$-wave interaction strengths for two spin states are identical is practically applicable. For example, Ref.\,\cite{lin1} shows that the difference of two-body interaction strengths for $^{87}$Rb in the spin-up state and the spin-down state is the order of $10^{-2}$ coupling length.
\end{thebibliography}
\end{document}